\documentclass{JFM-FLM_Au}

\lefttitle{Author 1 et al.}
\righttitle{Journal of Fluid Mechanics}

\title{Layering Theory of Liquids at Solid Interfaces: Interfacial Layering Oscillator Model}

\author{Chengzhen Sun \aff{1}, Yuntao Du \aff{1}, Tianyu Wu \aff{1}, and Mehdi Neek-Amal$^*$ \aff{2}}

\affiliation{
\aff{1} State Key Laboratory of Multiphase Flow in Power Engineering, Xi’an Jiaotong University, Xi’an, Shaanxi 710049, China \\
\aff{2} Departement Fysica, Universiteit Antwerpen, Groenenborgerlaan 171, B-2020 Antwerpen, Belgium\\

\aff{3} Department of Physics, Shahid Rajaee Teacher Training University, Lavizan, Tehran 16875-163, Iran}

\corresau{Mehdi Neek-Amal, mehdi.neek-amal@uantwerpen.be}

\begin{document}
\maketitle

\begin{abstract}
The structural organization of liquids near solid interfaces profoundly influences phenomena such as wettability, nanofluidic transport, and interfacial heat transfer. This study introduces the Interfacial Layering Oscillator Model (ILOM), a concise, semi-phenomenological framework that accurately captures the oscillatory density profiles of liquids adjacent to planar solid surfaces. By deriving a second-order differential equation rooted in classical statistical mechanics and calibrated with molecular dynamics simulations, ILOM predicts the amplitude, decay rate, and wavelength of interfacial density layering with exceptional computational efficiency. This versatile model applies to both hydrophilic and hydrophobic surfaces and extends to liquids beyond water, including methanol, providing valuable insights into critical interfacial properties that advance nanoscale fluid mechanics and material design.
\end{abstract}

\begin{keywords}
\end{keywords}

The microscopic behavior of liquids at solid interfaces governs a wide range of interfacial phenomena, including wetting, adhesion, slip, nanofluidic transport, and catalysis \citep{Bjorneholm2016,Geissler2013,PNAS2024}. At the molecular scale, liquids form structured layers near these interfaces, exhibiting oscillatory density profiles that decay over a few molecular diameters. These local density variations significantly impact macroscopic properties such as viscosity, thermal conductivity, and hydrodynamic boundary conditions \citep{barrat1999,hadjiconstantinou2015,neek2016commensurability}.

Recent theoretical and experimental studies have provided detailed insights into the layered structure of water at solid interfaces. Molecular dynamics simulations combined with density functional theory reveal that water molecules organize into hydration layers adjacent to both hydrophilic and hydrophobic surfaces, with layer spacings on the order of molecular diameters \citep{Milton2025}. Experimental techniques such as X-ray reflectivity and neutron scattering have substantiated these findings by resolving multiple discrete water layers with characteristic separations and density oscillations near the interfacial region \citep{AmannWinkel2016}. Furthermore, the interplay of hydrogen bonding networks, surface chemistry, and nanoscale confinement modulates these layers, affecting dynamic properties like diffusivity and interfacial friction \citep{Zhu2024}. Advancements in machine learning-based molecular simulation approaches further enhance the resolution and predictive capability regarding water’s interfacial orientation and structuring \citep{Zhu2024}. In a notable contribution, Wang and Hadjiconstantinou \citep{Wang_2017} developed the Nernst–Planck equation to model the fluid density profile at the fluid–solid interface, employing asymptotic analysis to derive polynomial expressions for the areal density of the first layer and demonstrating its importance in nanoscale transport phenomena. Collectively, these investigations underscore the fundamental role of interfacial layering in governing the physicochemical properties of water near solid surfaces, which is critical to a plethora of nanoscale processes.

Recent studies have shown that fluids confined within nanostructures such as carbon nanotubes or graphene slits exhibit densities significantly lower than those of the surrounding bulk reservoir \citep{Liu2005,Wang2004,Alexiadis2008}. This density anomaly has been directly linked to fluid layering \citep{Wang2015}, which also plays a crucial role in determining solvation pressures under confinement \citep{israelachvili2011intermolecular,Henderson1986}. The phenomenon of layering itself arises from interactions between the fluid and nearby solid atoms \citep{Rowley1976,Abraham1978,Thomas2008}, giving rise to distinct structural ordering at the boundary. Early investigations of this effect emphasized the structural features of adsorbed layers \citep{Thompson1990a} and highlighted their importance in tribological applications \citep{Thompson1990b}.

A particularly intriguing case is the behavior of water under extreme nanoconfinement in graphene slits, where the confinement width $w$ profoundly alters structure, dynamics, and pressure \citep{Pu2025,Fernandez2015,SobrinoFernandez2016,Jalali2020,Xu2024b,Jalali2021}. All-atom MD simulations reveal oscillatory translational and rotational dynamics correlated with layering, with hydration pressure varying from negative (hydrophobic attraction) to as high as $\sim 1$ GPa \citep{neek2016commensurability,Calero2020,AlgarSiller2015,Satarifard2017}. Free energy minima occur at widths commensurable with full layers (e.g., 7.0~\AA{} for monolayer, 9.5~\AA{} for bilayer), where pressure vanishes and dynamics slow, while incommensurable widths (e.g., $w \simeq 8.0$~\AA{}) lead to faster-than-bulk dynamics due to water expulsion \citep{Calero2020}. Continuum models incorporating microscopic structure explain enhanced flow rates via Hagen–Poiseuille with slip and disjoining pressure, attributing ultrafast flow to high density and low viscosity in ordered layers \citep{NeekAmal2018,Zhou2024}. Mean-field lattice models, such as Ono–Kondo variants, capture hysteresis in adsorption isotherms, linking low-pressure uptake to exponential dependence on adsorbate–adsorbent interactions and hydrogen bond networks in infinite slits \citep{Hawthorne2025}. These studies highlight how confinement induces phase transitions and anomalous transport, bridging MD insights with analytical predictions for applications like water purification \citep{Calero2020,Hawthorne2025}.

In particular, early statistical mechanical models, such as the Yvon–Born–Green (YBG) hierarchy \citep{henderson1992}, provide a rigorous yet computationally intensive description of density fluctuations in inhomogeneous fluids:
\[
\nabla \rho(\mathbf{r}) = -\beta \rho(\mathbf{r}) \int \nabla V(|\mathbf{r} - \mathbf{r}'|) \rho(\mathbf{r}') g(|\mathbf{r} - \mathbf{r}'|) \, d\mathbf{r}',
\]
where \(\beta = 1/(k_B T)\) and \(V\) denotes the pair potential (e.g., 12-6 Lennard-Jones). For a planar wall, this equation reduces to a one-dimensional integral equation for \(\rho(z)\), involving convolutions with the radial distribution function \(g(r)\), which balances forces from the wall and fluid-fluid interactions \citep{henderson1992}.

Similarly, density functional theory (DFT) and empirical wall potentials, such as Steele's 10-4-3 model \citep{steele1974, israelachvili2011intermolecular}, offer valuable insights into layering but often require system-specific tuning or significant computational resources. Molecular dynamics (MD) simulations provide detailed perspectives on interfacial dynamics, revealing how substrate properties influence the amplitude and decay of fluid density oscillations \citep{Wang_2017, Uhlig2021Atomically}. Recent studies have further highlighted the role of interfacial layering in enhancing thermal conductance at solid-liquid contacts, showing an exponential-to-linear crossover with interaction strength \citep{JPCC2024}. Moreover, machine learning-driven approaches have enabled high-throughput screening of interfacial structures, uncovering novel quantum phenomena at 2D material interfaces \citep{NSR2024Surface, NSR2025LiquidSolid}. However, advances in machine learning potentials, while improving accuracy, are limited by their substantial computational demands \citep{PNAS2024}.

Motivated by the need for a physically intuitive and computationally efficient approach, we propose a unified model that captures the essential physics of interfacial layering. By formulating a second-order differential equation grounded in classical theory and validated against MD simulations for water on both hydrophobic and hydrophilic surfaces, this model adapts to diverse interaction types and thermodynamic conditions. It offers predictive capabilities for interfacial properties—such as layering amplitude, decay length, and wavelength—serving as a valuable tool for theoretical and simulation-based studies of nanoscale transport.

\section{Model Formulation}

Liquids confined near solid surfaces exhibit oscillatory density profiles, a behavior driven by molecular packing and the interplay of fluid-solid and fluid-fluid interactions. We quantify this phenomenon by defining the normalized local density as \(g(z) = \rho(z) / \rho_{\text{bulk}}\), with the density deviation given by \(h(z) = g(z) - 1\).

Our approach builds on the theoretical foundation of the damped harmonic oscillator, drawing inspiration from the YBG equation, which describes structural correlations in inhomogeneous fluids. However, the YBG equation’s integro-differential form and dependence on complex pair correlations and nonlocal integrals make its direct solution computationally demanding. To address this, our ILOM is based on the development of a simplified, physics-based second-order ordinary differential equation (ODE) for \(h(z)\). This ODE effectively captures the key features of density layering—oscillatory patterns, exponential decay, and confinement effects—while ensuring computational tractability. A detailed derivation linking this ODE to the YBG framework is provided in the Appendix A.

To construct a simplified model for the density layering of liquids near solid interfaces, we begin with a phenomenological motivation but also aim to connect it with established theoretical foundations. To this end, we start from the YBG integral equation and transform it into a more tractable form. As shown in Appendix~A, by straightforward differentiation, the YBG equation is converted into an integro-differential form that incorporates both wall--fluid and fluid--fluid interactions. Although this formulation provides a rigorous description, the nonlocal integral terms introduce substantial computational complexity.  

To simplify the treatment, we linearize these terms by assuming small deviations from the bulk density, specifically \( h(z') \approx 0 \) for \( z' > \sigma_{\text{ll}} \), where \( \sigma_{\text{ll}} \) represents the molecular length scale. This approximation emphasizes the dominant short-range contributions, enabling a logical reduction of the problem to its essential physical ingredients. The outcome is a manageable differential equation that retains the fundamental characteristics of the system while remaining computationally feasible.  

By applying these approximations (detailed in Appendix~A) and accounting for both wall and fluid--fluid effects, we obtain a preliminary differential equation unifying the two interaction types:
\begin{equation}
\frac{d^2 h(z)}{dz^2} + \beta \left[ \frac{d h(z)}{dz} \left( \frac{d V_{\text{wall}}}{dz} + \rho_{\text{bulk}} \frac{\partial V_{ll,\text{eff}}}{\partial z} e^{-z / \sigma_{\text{ll}}} \right) + (h(z) + 1) \frac{d^2 V_{\text{wall}}}{dz^2} \right] \approx 0.
\label{eq:3}
\end{equation}
This equation represents a simplified version of the full expression in Eq.~(\eqref{eq:A2}). The first term on the right-hand side reflects the combined damping effect from wall and fluid--fluid forces, while the second term corresponds to the restoring force arising from the curvature of the wall potential. The exponential decay factor ensures that fluid--fluid interactions vanish at larger distances, preserving the physical behavior of the layering profile.  

To extract physical insight, Eq.~\eqref{eq:3} is rearranged to isolate key contributions. The damping term, associated with the velocity-like quantity \( \frac{dh}{dz} \), captures dissipative effects due to interfacial interactions, while the oscillatory term, linked to the curvature of the potential, represents the restoring force responsible for periodic layering. Together, these terms naturally lead to the form of a damped harmonic oscillator, providing a transparent mathematical framework for describing the oscillatory density profile and its exponential decay into the bulk fluid.  

Consequently, the final governing equation can be written as an excited, damped one-dimensional harmonic oscillator:
\begin{equation}
\frac{d^2 h}{dz^2} + 2 \gamma(z) \frac{dh}{dz} + \omega^2(z) h = F(z),
\label{eq:4}
\end{equation}
where each term has a clear physical meaning:
\begin{itemize}
    \item \textbf{Inertial term} (\(\frac{d^2 h}{dz^2}\)): Represents the curvature of the density profile, capturing rapid spatial variations near the wall.
    \item \textbf{Damping term} (\(2 \gamma(z) \frac{dh}{dz}\)): Describes energy dissipation due to wall--fluid and fluid--fluid interactions.
    \item \textbf{Oscillatory term} (\(\omega^2(z) h\)): Governs the periodic structuring of molecules driven by the potential landscape.
    \item \textbf{Forcing term} (\(F(z)\)): Represents density modulations arising from fluid--fluid correlations.
\end{itemize}

This systematic derivation provides a bridge between microscopic statistical mechanics and a phenomenological continuum description of layering. Although Eq.~\eqref{eq:4} can also be introduced empirically, its approximate derivation from the YBG framework offers a physically grounded foundation for understanding interfacial density oscillations.

Here, we introduce three model variants based on the treatment of \(\gamma(z)\), \(\omega(z)\), and \(F(z)\):
\begin{itemize}
    \item The simplest, termed the ``Simple Harmonic Oscillator (SHO)'' model, assumes constant \(\gamma_0\) and \(\omega_0\), with \(F(z) = 0\). The solution to Eq.    \eqref{eq:4} is:
      \begin{equation}
      h(z) = h_0 e^{-\gamma_0 z} \cos(\omega_d z + \phi),
      \label{eq:5}
      \end{equation}
      where \(\omega_d^2 = \omega_0^2 - \gamma_0^2\). 
      Consequently, \(\rho(z) = (h(z) + 1) \rho_{\text{bulk}}\). In an ideal case, density peaks occur when \(\cos(\omega_d z_n + \phi) = 1\), at positions \(z_n = \frac{2\pi n}{\omega_d} - \frac{\phi}{\omega_d}\) (for \(n \geq 1\)). This can be visualized as a strobe light highlighting the oscillator’s maxima, with amplitude \(h_n = h_0 e^{-\gamma_0 z_n}\) decaying exponentially due to damping. Note that these \(z_n\) are not exact for a real liquid at the interface.
    \item The ``SHO1'' model extends this by allowing \(\gamma(z)\) and \(\omega(z)\) to vary with \(z\), while keeping \(F(z) = 0\). The density profile \(g(z) = \rho(z) / \rho_{\text{bulk}}\) is computed by solving the ODE for \(h(z)\) as a boundary value problem (BVP). Parameters are optimized against MD data using a Python implementation, with further details in the Appendix B.
    \item The ``SHO2'' model incorporates a non-zero \(F(z)\) to account for additional correlation effects, though this variant is not explored further in this study.
\end{itemize}

For the SHO1 and SHO2 models, the position-dependent damping coefficient is approximated as
\begin{equation}
\gamma(z) = \gamma_1 \left(1 + \alpha e^{-z / \sigma_{\text{ll}}} \right),
\end{equation}
where \(\gamma_1\) is the bulk damping coefficient, \(\alpha\) controls the near-wall enhancement, and \(\sigma_{\text{ll}} \approx 3~\text{\AA}\) represents the characteristic molecular length scale of the fluid.  

The oscillatory term is modeled as
\begin{equation}
\omega^2(z) = \omega_1^2 e^{-\lambda z / \sigma_{\text{ll}}} + \omega_{\text{bulk}}^2,
\end{equation}
where \(\omega_1^2\) determines the near-wall packing frequency, \(\omega_{\text{bulk}}^2\) corresponds to the bulk oscillation frequency, and \(\lambda\) regulates the exponential decay of wall-induced effects.  

For the SHO2 model, the external forcing term is given by
\begin{equation}
F(z) = -\kappa e^{-z / \sigma_{\text{ll}}},
\end{equation}
where \(\kappa\) scales with the squared average density, i.e., \(\kappa \propto \rho_{\text{ave}}^2\).  
This functional form ensures that wall-induced forcing diminishes smoothly into the bulk, consistent with the decay of density modulation observed in molecular simulations.

The numerical ODE becomes:
\begin{equation}
\frac{d^2 h}{dz^2} + 2 {\gamma_1} \left(1 + \alpha e^{-z / \sigma_{\text{ll}}} \right) \frac{dh}{dz} + \left[ \omega_1^2 e^{-\lambda z / \sigma_{\text{ll}}} + \omega_{\text{bulk}}^2 \right] h = -\kappa e^{-z / \sigma_{\text{ll}}},
\label{eq:9}
\end{equation}
satisfying boundary conditions from MD: \(h(z \to \infty) \to 0\) and a finite derivative at \(z = 0\). The optimization is performed iteratively, minimizing the mean squared error (MSE) between model predictions and MD-derived $ g(z) $, ensuring physical consistency across diverse substrates and thermodynamic conditions; see the Appendix B for details.

\begin{figure}[h]
    \centering
    \includegraphics[width=0.7\linewidth]{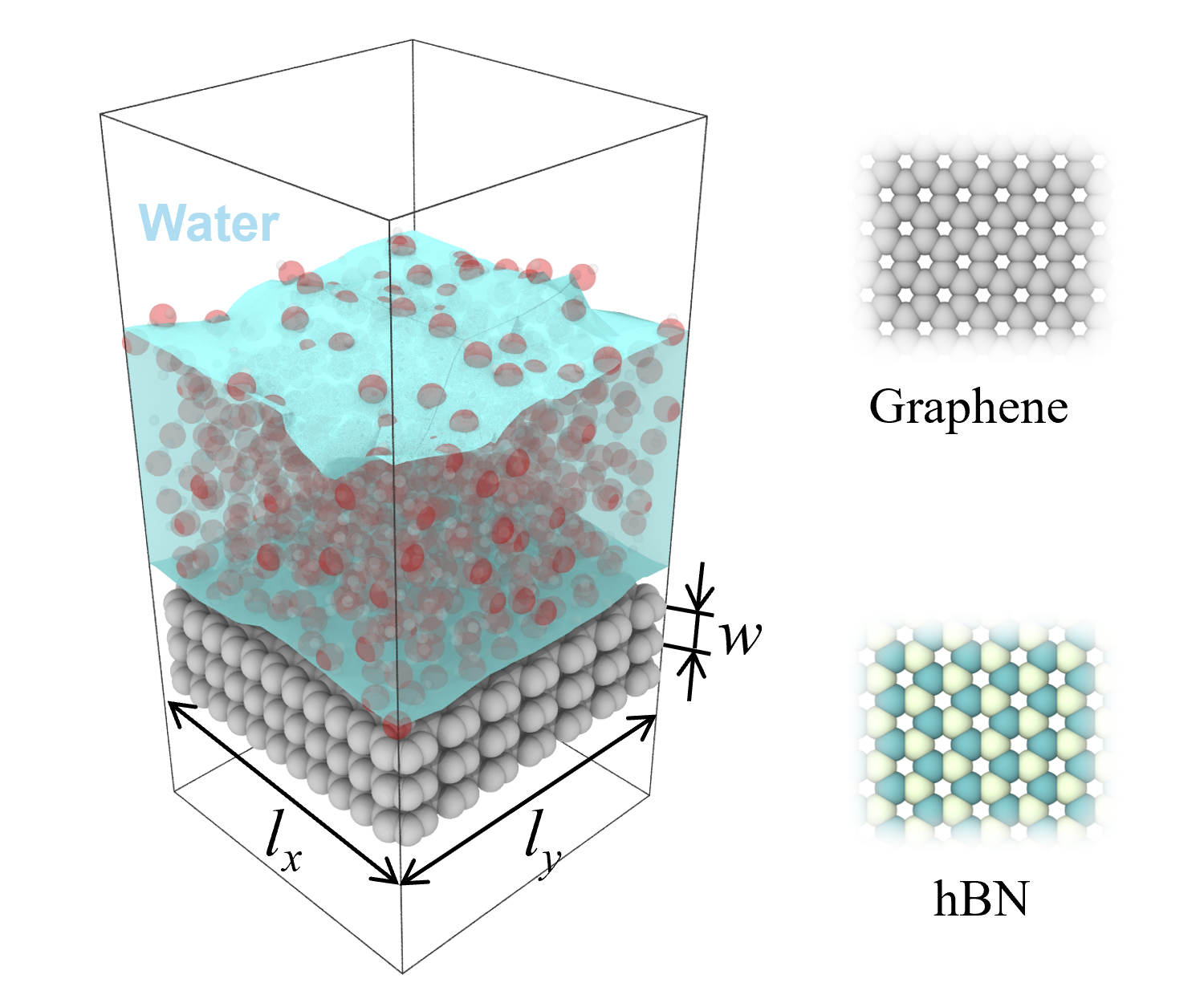}
    \caption{Illustration of the MD model configurations for water over graphene and hexagonal boron nitride (hBN) surface.}
    \label{fig:md_model}
\end{figure}

\section{Molecular Dynamics Simulation Details}

Equilibrium MD simulations were performed using the LAMMPS package \citep{Thompson2022,Plimpton1995} to investigate the layering behavior of water reservoirs on flexible graphene and rigid hexagonal boron nitride (hBN) surfaces. Water interatomic interactions were modeled using the 12-6 Lennard-Jones (LJ) potential combined with Coulomb interactions, employing the SPC/E water model. The SHAKE algorithm ensured the maintenance of constant bond lengths and angles in water molecules.

Potential parameters, sourced from \cite{Sun2024, Hilder2024}, are listed in table~\ref{tab:params}. Interactions between different atom types followed the Berthelot-Lorentz mixing rules \citep{Lorentz1881,Berthelot1898}. Long-range electrostatic interactions were handled using the Particle-Particle-Particle-Mesh (PPPM) method, with a tolerance of \(10^{-4}\). All cutoff distances were set to \(10.0 \, \text{Å}\), and the simulation timestep was \(1 \, \text{fs}\).

Simulations were conducted in the NVT ensemble, with the temperature maintained at \(300 \, \text{K}\) using a Nosé-Hoover thermostat. Each system was equilibrated for \(1 \, \text{ns}\), followed by a \(3 \, \text{ns}\) data collection phase with a sampling interval of \(1 \, \text{ps}\).

\begin{table}
    \centering
    \begin{tabular}{lcccc}
        Molecule & Site & \(\varepsilon\) (kcal/mol) & \(\sigma\) (\AA) & \(q\) (e) \\[3pt]
        H\(_2\)O & O & 0.1554 & 3.1660 & -0.8476 \\
                 & H & 0.0000 & 0.0000 & 0.4238 \\
        Methanol & [CH$_3$]-O-H & 0.1950 & 3.7500 & 0.265 \\
                 & CH$_x$-[O]-H & 0.1850 & 3.0200 & -0.700 \\
                 & O-[H] & 0.0000 & 0.0000 & 0.4350 \\                 
        Graphene & C & 0.0556 & 3.4000 & 0.0000 \\
        hBN & N & 0.1450 & 3.3650 & -0.9300 \\
            & B & 0.0949 & 3.4530 & 0.9300 \\        
    \end{tabular}
    \caption{Potential parameters used in the MD simulations.}
    \label{tab:params}
\end{table}

MD simulations were conducted to study water reservoirs on flexible graphene (hydrophobic) and rigid hBN (hydrophilic) surfaces, as illustrated in figure ~\ref{fig:md_model}. Periodic boundary conditions were applied along the \(x\) and \(y\) directions, with fixed boundaries in the \(z\) direction. Each reservoir contained 640 water molecules, resulting in a relaxed thickness of approximately \(20 \, \text{Å}\). The simulation box for graphene measured \(l_x \times l_y = 29.69 \times 29.40 \, \text{Å}^2\), with the structure of both cases comprising three layers spaced by \(w = 3.40 \, \text{Å}\). The top layer of carbon atoms in the flexible graphene surface was free to move, while the underlying layers remained fixed, and a Langevin thermostat maintained the graphene temperature at \(300 \, \text{K}\).

\section{Results and Discussion}\label{sec:results}
\subsection{Water over graphene and hBN}
\begin{figure*}[h]
    \centering
    \includegraphics[width=1\linewidth]{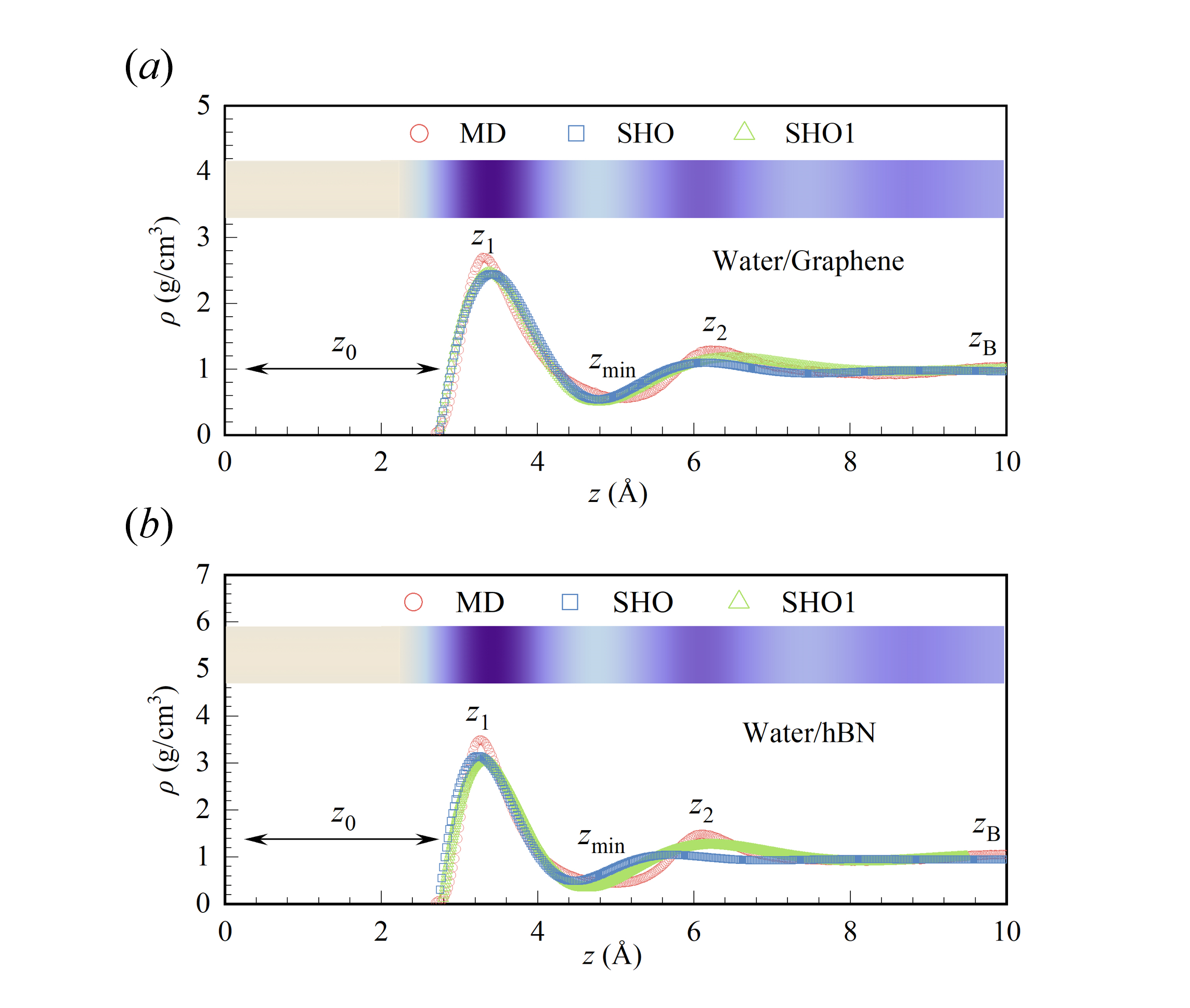}
    \caption{Density profiles of water near graphene (a) and hBN (b) surfaces, each exhibiting three distinct layers. Blue square dots represent the SHO model Eq.\eqref{eq:5}, green triangular dots the SHO1 model, and red circle dots MD simulation data for comparison. The color bar indicates the density distribution across the channel, measured from the graphene wall.}
    \label{fig:water_graphene}
\end{figure*}

We first present results from the SHO model for water over graphene and hBN. MD data, alongside the best fits to Eq.\eqref{eq:5}, are shown as red and blue dots in figure~\ref{fig:water_graphene}. The optimized parameters \(\gamma_0\), \(h_0\), \(\omega_d\), \(\phi\), and \(h_s\) are listed in table~\ref{tab:SHO_parameters}.

\begin{table}
    \centering
    \begin{tabular}{cccccc}
        Surface & \(h_0\) & \(\gamma_0\) (\AA$^{-1}$) & \(\omega_d\) (\AA$^{-1}$) & \(\phi\) & \(\rho_{bulk}\) \\[3pt]
        water/Graphene & 3.1105243234 & 0.896693164 & 2.294043769 & \(1.356 \pi\) & 0.969478004 \\
        water/hBN & 5.2157063399 & 1.266575162 & 2.543868249 & \(1.421\pi\) & 0.944793255 \\
            methanol/graphene & 3.30583559441 &0.521760822 & 1.993393789 & \(1.4055\pi\) & 0.769508397 \\
  
    \end{tabular}
    
    \caption{Optimized parameters for the SHO model for water over graphene and hBN as well as methanol over graphene. For each case, the bulk density $\rho_{\text{bulk}}$ is determined from the corresponding MD simulations in the limit of large $z$ values.}
    \label{tab:SHO_parameters}    
\end{table}

As expected, \(h_0\) approximates the density at the first peak \(\rho(z_1)\), representing the initial amplitude. The damping coefficient \(\gamma_0\), reflecting energy dissipation, while \(\omega_d \) corresponds to the wavelength of density oscillations, influenced by molecular spacing and confinement. The first peak represents a strongly adsorbed layer (akin to the Stern layer in electrochemistry), where water molecules are tightly bound to the surface via van der Waals or electrostatic forces, leading to higher density. Subsequent layers resemble the diffuse layer, with oscillations decaying as intermolecular forces weaken with distance.

\begin{table}
    \centering
    \begin{tabular}{ccccccc}
      Surface & \(\gamma_1\) (\(\sigma^{-1}\)) & \(\alpha\) & \(\omega_1^2\) (\(\sigma^{-2}\)) & \(\omega_{\text{bulk}}^2\) (\(\sigma^{-2}\)) & \(\lambda\) \\[3pt]
        water/Graphene & 0.339734605 & 5.02229170 & 84.6312461 & 4.09846193 & 3.44037088 \\
        water/hBN & 0.218736893 & 10.9428934 & 21.9911135 & 0.962598384 & 1.14422473 \\
          methanol/Graphene& 0.291260941 & 1.77863091 & 20.8553699 &3.03057967& 2.21661477 \\
    \end{tabular}
    \caption{Optimized parameters for the SHO1 model for water over  graphene and hBN as well as methanol over graphene.}
    \label{tab:optimized_parameters}    
\end{table}

Notably, $z_0$, representing the first non-zero value of the $\rho(z)$, is approximated as $(2/5)^{1/6} \sigma$, where the wall-fluid interaction is zero, approximately 2.83 \AA ~for water over graphene (see below for details on wall-fluid interactions). Consequently, all fits in the plots begin at the corresponding $z_0$ value.

The best-fit parameters for graphene (hBN), determined using the Levenberg-Marquardt algorithm with least squares over nine iterations, show robust agreement with MD data. Statistical metrics include an \(r^2\) of 0.924 (0.895), an adjusted \(r^2\) of 0.923 (0.894), a fit standard error of 0.139 (0.217), and a maximum absolute error of 0.384 (0.472), indicating excellent concordance, particularly around the first peak and its width.

To enhance the fit around the first minimum (\(z_{\text{min}}\)) and second peak (\(z_2\)) for water over graphene and hBN, we applied the SHO1 model with parameters from table~\ref{tab:optimized_parameters}. This model minimizes the mean squared error (MSE) between \(g(z)\) and MD data using a differential evolution algorithm with a population size of 40 and up to 100 iterations. If convergence fails, an error flag triggers a large MSE penalty (see Appendix B for details).

Analysis of figure~\ref{fig:water_graphene} (green  dots) reveals that the SHO1 model preserves the first peak but improves the fit at the first minimum and second peak, attributable to the position-dependent \(\gamma(z)\) and \(\omega(z)\). The enhancement is more pronounced for hBN around the first minimum and second peak compared to graphene. Further parameter refinement could yield closer agreement, though the current concordance between MD and SHO models is already highly encouraging, reducing the need for extensive optimization in this study.

\begin{figure*}
    \centering
    \includegraphics[width=1\linewidth]{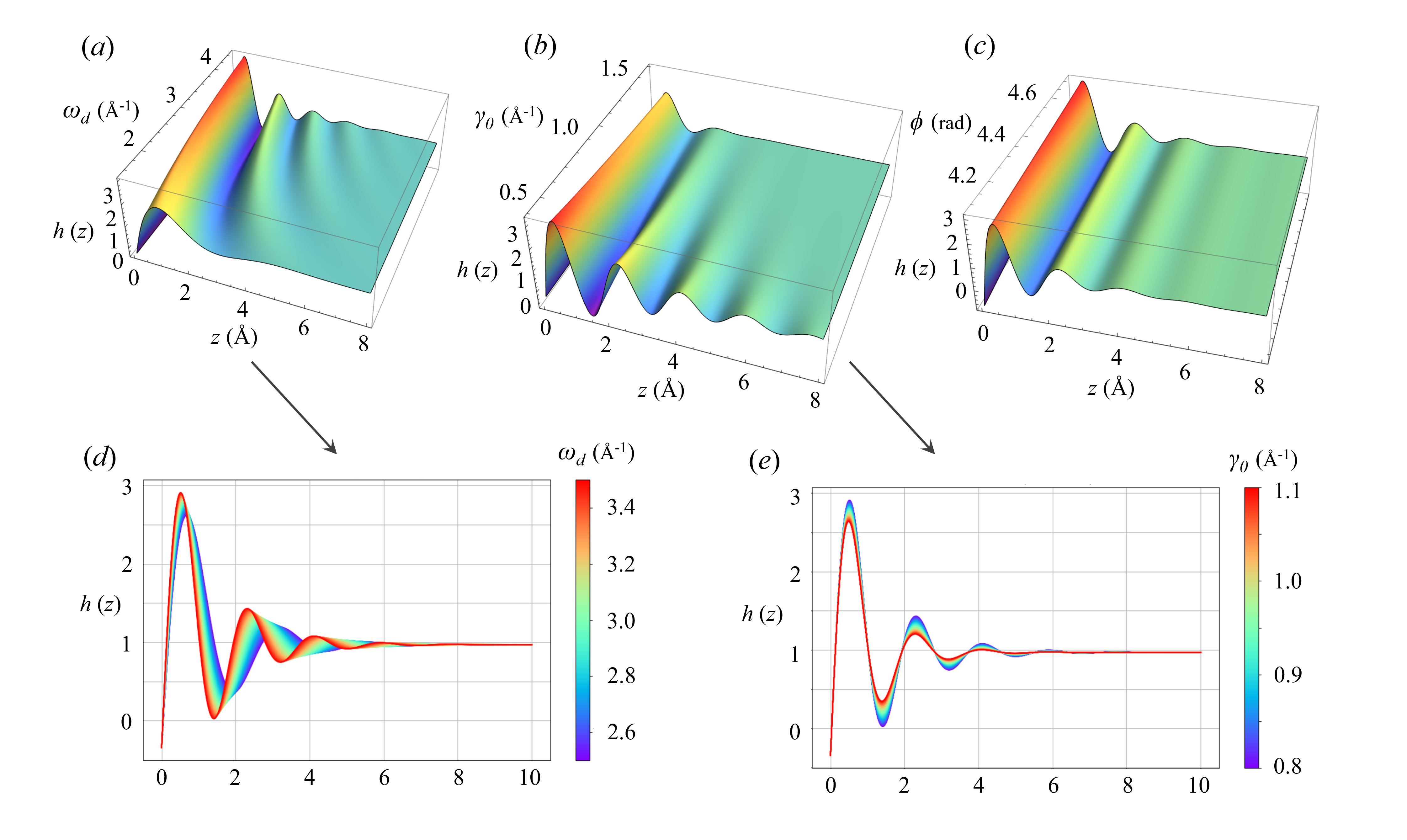}
    \caption{Density plot of the density variation according to the SHO solution Eq.\eqref{eq:5} for (a) \(\gamma_0 = 0.80 \, \text{Å}^{-1}\), \(\phi = 1.356\pi\), and \(h_s = 0.97\), with \(\omega_d\) ranging from 1.5 to 4.5 \(\text{Å}^{-1}\) and \(z - z_0\) from 0 to 8 \(\text{Å}\); and (b) \(\omega_d = 3.5 \, \text{Å}^{-1}\), \(\phi = 1.356\pi\), and \(h_s = 0.97\), with \(\gamma_0\) ranging from 0.4 to 1.5 \(\text{Å}^{-1}\) and \(z - z_0\) from 0 to 8 \(\text{Å}\). Panel (c) shows the effects of \(\phi\) for \(\gamma_0 = 0.80 \, \text{Å}^{-1}\), \(h_s = 0.97\), and \(\omega_d = 3.5 \, \text{Å}^{-1}\). The bottom panels (d) and (e) present the front view of (a) and (b) for a narrower range of $\omega_d$ and $\gamma_0$ respectively.}
    \label{fig:2D}
\end{figure*}

To illustrate the effects of \(\omega_d\) and \(\gamma_0\) in figure~\ref{fig:2D}, we present density plots showing the variation of \(h(z)\) with \(\omega_d\) and \(z\) in panel (a) for \(\gamma_0 = 0.80 \, \text{Å}^{-1}\), \(\phi = 1.356\pi\), and \(h_s = 0.97\), where \(\omega_d\) ranges from 1.5 to 4.5 \(\text{Å}^{-1}\) and \(z\) from 0 to 8 \(\text{Å}\). Panel (b) shows the variation of \(h(z)\) with \(\gamma_0\) and \(z\) for \(\omega_d = 3.5 \, \text{Å}^{-1}\), \(\phi = 1.356\pi\), and \(h_s = 0.97\), where \(\gamma_0\) ranges from 0.4 to 1.5 \(\text{Å}^{-1}\) and \(z\) from 0 to 8 \(\text{Å}\). Increasing \(\omega_d\) or decreasing \(\gamma_0\) enhances the oscillatory profile, mimicking the layered structure associated with stronger adsorption at interfaces or the structure of an LJ liquid, as discussed below. Panel (c) demonstrates that the effect of \(\phi\) is minimal for \(\phi\) between \(1.3\pi\) and \(1.5\pi\). 

\begin{figure*}[!htb]
    \centering
    \includegraphics[width=1\linewidth]{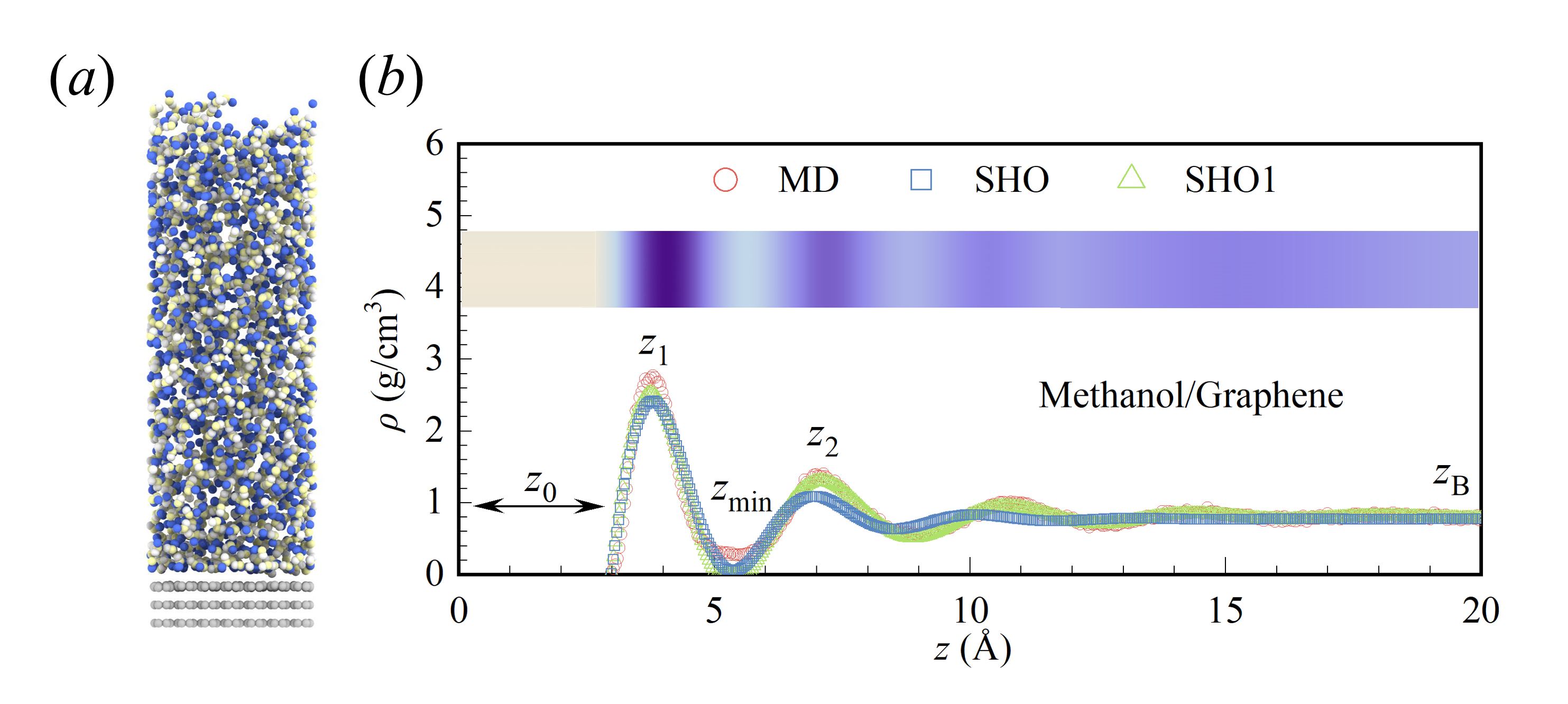}
    \caption{(a) Illustration of the molecular dynamics (MD) model configurations for methanol over graphene. (b) Density profiles of methanol near graphene  surfaces. Blue square dots represent the SHO model Eq.\eqref{eq:5}, green triangular dots the SHO1 model, and red circle dots MD simulation data for comparison. The color bar indicates the density distribution across the channel, measured from the graphene wall.}
    \label{figmethanol}
\end{figure*}

\subsection{Methanol over graphene}

To showcase the robustness and versatility of our modeling framework for a broader range of liquids beyond water, we conducted comprehensive MD simulations for methanol (CH$_3$OH). The simulated systems, featuring methanol reservoirs on a flexible graphene surface, are vividly illustrated in figure.~\ref{figmethanol}(a). The graphene surface configuration remains consistent with our prior setup, with the sole distinction being the substitution of water with methanol. For these simulations, we employed the TraPPE-UA united-atom force field, meticulously optimized for alcohol-based organic molecules, including both methanol and ethanol \citep{chen2001}. Interatomic interactions are skillfully handled using the Berthelot-Lorentz mixing rule, ensuring accurate representation of the molecular interactions. The resulting MD data for methanol density profiles are elegantly displayed in figure.~\ref{figmethanol}(b). These profiles were derived by computing the centroid of each methanol molecule, revealing a commendable alignment with the bulk density values as the liquid extends far from the surface.

To interpret these density profiles, we applied the SHO and SHO1 models, with their fits represented by blue and green dots in figure.~\ref{figmethanol}(b). Both models excel in precisely identifying the positions of the first peak and first minimum. However, the depth of the first minimum poses a challenge, as it is not fully replicated by either model. Remarkably, the SHO1 model stands out by delivering exceptionally accurate depictions of the second and third peaks, offering an outstanding fit across regions both beyond and prior to $z_{\text{min}}$. The detailed fitting parameters for the SHO and SHO1 models are meticulously cataloged in tables 2 and 3, respectively, providing a clear reference for their performance.

A critical observation concerns the molecular structure of methanol, which consists of six atoms and exhibits significant asymmetry due to the contrasting methyl group (CH$_3$) and hydroxyl group (OH), which differ markedly in size and polarity. This structural complexity contrasts sharply with the more symmetric and simpler three-atom structure of water (H$_2$O). These differences notably affect the depth and definition of the first minimum in the density profile. Consequently, the SHO and SHO1 models, though highly effective, may not fully capture the depth of this minimum, highlighting the inherent challenges posed by methanol's structural asymmetry and the presence of two heavily weighted carbon (C) and oxygen (O) atoms in its structure, which adsorb more strongly to the surface and result in a deeper first minimum compared to water with only one such atom—fluctuations that are considerably less pronounced in the more isotropic water molecule.

\subsection{Lennard-Jones liquid over graphene}
Finally, we evaluated the SHO1 and SHO2 models with an artificial LJ liquid over graphene, inspired by Wang et al. \citep{Wang_2017}. The LJ potential is:
\begin{equation}
V(r) = 4 \varepsilon \left[ \left( \frac{\sigma_f}{r} \right)^{12} - \left( \frac{\sigma_f}{r} \right)^6 \right],
\end{equation}
with \(\varepsilon = 0.15 \, \text{kcal/mol}\) and \(\sigma_f = 3.15 \, \text{Å}\) for fluid-wall interactions, and identical \(\sigma_f\) and \(\varepsilon_f\) for fluid-fluid interactions, with a \(4 \sigma_f\) cutoff \citep{Wang_2017}.

\begin{figure}[!htb]
    \centering
    \includegraphics[width=0.75\linewidth, height=0.35\textheight]{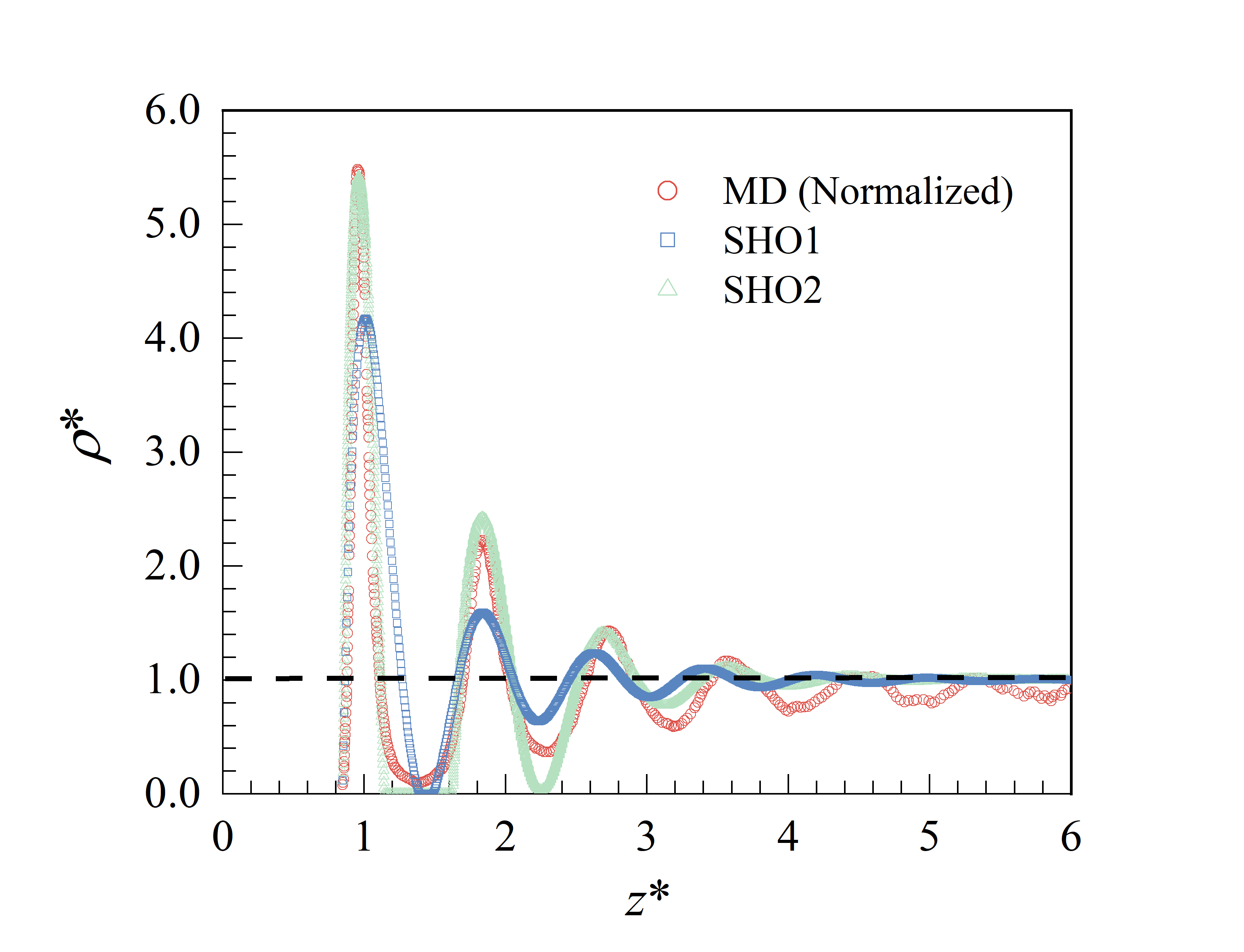}
    \caption{Normalized density profiles for a LJ liquid over graphene. Blue square dots represent the SHO1 model Eq.\eqref{eq:9}, {{green triangular dots the SHO2 model }}, and red circle dots MD simulation data for comparison. Optimized parameters from table~\ref{tab:params_LJ} align well with the first two peaks.}
    \label{fig:S2}
\end{figure}

MD simulations using LAMMPS in the NVT ensemble, with a Nosé-Hoover thermostat, employed a \(6.25 \times 10^{-4}\) time step and 4-time-unit equilibration to mimic graphene surface conditions. Optimized parameters are listed in table~\ref{tab:params_LJ}, with additional profiles in figure.~\ref{fig:S2}). The model accurately reproduces the first three peak positions, widths, and amplitudes but underperforms at the first minimum compared to water over graphene. This strong agreement validates its ability to capture initial layering due to intense wall-fluid interactions and predict layer spacing and amplitude decay, supporting studies of nanofluidic transport.  In Fig. \ref{fig:S2}, $ z^* = z / \sigma_{f} $ and $ \rho^* = \rho \sigma_{f}^3 $, where $ z $ is the distance from the interface, and $ \rho $ is the number density of the fluid.

\begin{table}
    \centering
    \begin{tabular}{ccccccc}
        & \(\gamma_1\) (\(\sigma^{-1}\)) & \(\alpha\) & \(\omega_1^2\) (\(\sigma^{-2}\)) & \(\omega_{\text{bulk}}^2\) (\(\sigma^{-2}\)) & \(\lambda\) & \(\kappa\) (\(\sigma^{-1}\)) \\[3pt]
        SHO1 & 1.08 & 50.49 & 1.29 & 65.22 & 2.94 & 0.0 \\
        SHO2 & 1.57 & 6.08 & 0.71 & 54.59 & 2.90 & 3482.51 \\
    \end{tabular}
    \caption{Optimized parameters for the LJ liquid system over graphene using SHO1 and SHO2 models.}
    \label{tab:params_LJ}
\end{table}

The SHO model's effectiveness stems from its representation of the liquid's oscillatory density profile as a damped harmonic oscillator, reflecting the physical interplay of molecular packing, fluid-solid interactions, and energy dissipation near the solid interface. The oscillations and exponential decay in the density profile are closely linked to the layered structure of water, where stronger adsorption at the first layer—due to enhanced van der Waals and hydrogen-bonding interactions with the hBN surface—produces a pronounced initial peak. Subsequent layers exhibit diminishing amplitudes as intermolecular forces weaken with distance, resembling the Stern layer in electrochemistry, where a compact ion layer forms near a charged surface before transitioning to a diffuse region. The decay length, influenced by the Debye length in charged systems, correlates here with the molecular length scale \(\sigma_{\text{ll}}\), modulating the screening of wall effects. This model's success in capturing these features underscores its potential to bridge classical statistical mechanics with nanoscale interfacial phenomena, providing insights into confinement-induced structuring across diverse systems.

\subsection{Discussion on Fluid–Solid Interface Structure}

The structural organization of liquids at solid interfaces has been widely studied. Wang and Hadjiconstantinou \citep{Wang_2017} showed that near solid surfaces, fluid density profiles exhibit molecular-scale oscillations that strongly affect nanoscale heat and momentum transfer. Using molecular mechanics and MD simulations, they analyzed the first fluid layer in the so-called layering regime, governed by the ratio of wall–fluid interaction strength to thermal energy.
This framework complements our simple harmonic oscillator (SHO) model and provides physical grounding for the interpretation of layering phenomena.

We propose heuristic relationships for the SHO model parameters \( \omega_d \), \( \gamma_0 \), \( h_0 \), and \( \phi \) based on the LJ interaction parameters \( \sigma \) and \( \varepsilon \), derived from the wall potential,
\[
V_{\text{wall}}(z) = 2\pi \varepsilon \rho_s \sigma^2 
\left[ \frac{2}{5}\left(\frac{\sigma}{z}\right)^{10} - \left(\frac{\sigma}{z}\right)^4 \right],
\]
where \( \varepsilon \) is the energy scale and \( \rho_s \) is the solid areal density. These relationships are constructed using dimensional analysis, physical reasoning, and a semi-empirical framework, as a full analytical derivation would require detailed MD simulations or advanced statistical mechanical treatments. Our semi-empirical approach integrates the damped harmonic oscillator model, the LJ wall potential, and insights from \cite{Wang_2017}. The derivations maintain dimensional consistency, rely on scaling arguments, and reflect the structure of the density profile near the interface. The resulting expressions represent approximation influenced solely by van der Waals interactions, subject to further validation against numerical or experimental data.

The SHO model solution, Eq.~\eqref{eq:5}, reads
\[
h(z) = h_0 e^{-\gamma_0 z} \cos(\omega_d z + \phi) + h_s,
\]
where \( \omega_d = \sqrt{\omega_0^2 - \gamma_0^2} \). The density modulation \( h(z) \) arises from the balance between wall–fluid interactions (via \( V_{\text{wall}} \)) and fluid–fluid correlations, as captured by the corresponding ODE (Eq.~\ref{eq:4}). We focus on dense fluids (\( \rho_{\text{ave}} \geq 0.7 \)) where pronounced layering occurs and thermal energy \( k_B T \) competes with wall attraction energy \( \varepsilon \).

\subsubsection{$ \omega_d \propto \sqrt{\frac{\varepsilon \rho_s}{k_B T}} $}
$ \omega_d $ governs the oscillation wavelength, tied to molecular spacing influenced by $ \sigma $. The frequency increases with stronger wall interactions ($ \varepsilon \rho_s $) relative to thermal energy ($ k_B T $), enhancing layering. Our estimation, grounded in van der Waals interactions, provides a lower limit for $ \omega_d $, with fitted values accounting for additional fluid-fluid contributions. $ \omega_d $ has units of length$^{-1}$. The term $ \frac{\varepsilon \rho_s}{k_B T} $ has units of length$^{-2}$, and the square root ensures the correct length scaling.

The oscillation frequency $ \omega_0 $ (before damping) is derived from the second derivative of $ V_{\text{wall}} $ at its minimum ($ z = \sigma $, where 

{\(  \frac{d V_{\text{wall}}}{dz} = 0 $)}:
$ \frac{d^2 V_{\text{wall}}}{dz^2} ={ 2\pi \varepsilon \rho_s} \left[ 44 \left( \frac{\sigma}{z} \right)^{12} - 20 \left( \frac{\sigma}{z} \right)^6 \right]. \)

At $ z = \sigma $, this simplifies to $ 48\pi \varepsilon \rho_s $. In the ODE, $ \omega^2 \sim \frac{1}{k_B T} \frac{d^2 V_{\text{wall}}}{dz^2} $, so $ \omega_0 \propto \sqrt{\frac{\varepsilon \rho_s}{k_B T}} $. For weak damping, $ \omega_d \approx \omega_0 $.

{\( \gamma_0 \propto \sqrt{\frac{\varepsilon \rho_s}{k_B T}} \)} \( \gamma_0 \) represents damping, dependent on interaction strength (\( \varepsilon \rho_s \)). Our estimation offers a lower limit, with fitted values reflecting additional damping from fluid dynamics. \( \gamma_0 \) has units of length$^{-1}$. The term \( \frac{\varepsilon \rho_s}{k_B T} \) has units of length$^{-2}$, and the square root provides the correct scaling. Damping aligns with the decay length of oscillations, consistent with MD and DFT \citep{evans1979, Wang_2017}, enhanced by stronger wall interactions. 

{\( h_0 \propto \frac{\varepsilon \rho_s \sigma^2}{k_B T} \)} \( h_0 \) is the initial amplitude of the density deviation, driven by first-layer adsorption strength (\( \varepsilon \rho_s \)) and normalized by \( \sigma^2 \) (area per molecule) and \( k_B T \) (thermal energy). The estimated \( h_0 \) represents a lower limit based on wall interactions, with fitted values indicating enhanced amplitudes due to fluid structuring. \( h_0 \) is dimensionless. \( \varepsilon \rho_s \sigma^2 \) has units of energy, matching \( k_B T \). The areal density \( \Sigma_{\text{FL}} \) (from \cite{Wang_2017}) scales as \( \rho_{\text{ave}} h_{\text{FL}} \), and \( h_0 \sim \frac{\Sigma_{\text{FL}}}{\rho_{\text{bulk}}} \). The ratio \( \frac{\varepsilon \rho_s \sigma^2}{k_B T} \) reflects adsorption energy relative to thermal energy. 

{\( \phi \propto \arctan\left(\frac{\sigma}{z_0}\right) \)} \( \phi \) shifts the oscillations, determined by the position of the potential minimum relative to the wall. \( \sigma \) sets the length scale, and \( z_0 \) is a reference distance. The estimated \( \phi \) provides a lower limit based on wall potential geometry, with fitted values adjusted by additional phase effects. \( \phi \) is dimensionless, and \( \frac{\sigma}{z_0} \) is dimensionless. \( \arctan \) ensures a bounded shift. The minimum of \( V_{\text{wall}} \) at \( z = \sigma \) aligns the peak, suggesting \( \phi \propto \arctan(\sigma / z_0) \).

These relations are dimensionally consistent but require validation with MD simulations or ODE solutions using the full LJ potential.

For \( \varepsilon \approx 0.08~\text{kcal/mol} \), 
\( \rho_s = 0.38~\text{Å}^{-2} \), and \( k_B T \approx 0.6~\text{kcal/mol} \) (300~K), 
we estimate
\[
\gamma_0 = \sqrt{\frac{\varepsilon \rho_s}{k_B T}} \approx 0.225~\text{Å}^{-1},
\quad
\omega_d \approx 2.76~\text{Å}^{-1}, \quad
h_0 \approx 0.456, \quad
\phi \approx 0.25\pi
\] which are close to the fitting parameters reported in table \ref{tab:SHO_parameters}. 
Other values are summarized in tables~\ref{tab:S1}. To assist readers in understanding the meaning of each parameter used in the paper, we have provided two comprehensive tables, Tables \ref{tab:parameters_ilom} and \ref{tab:parameters_lj_md}, summarizing the parameters of the Interfacial Layering Oscillator Model (ILOM) and the Lennard-Jones (LJ) potential and molecular dynamics (MD) simulations, respectively.

\section{Conclusion}

We have developed the Interfacial Layering Oscillator Model (ILOM), a compact, physically grounded framework to predict the density layering of fluids, particularly water, near solid interfaces. Encapsulated in a second-order differential equation, ILOM synthesizes insights from classical and contemporary methodologies, offering predictive capabilities across diverse substrates and thermodynamic conditions. Notably, our model accurately captures the first maximum, width, first minimum, and other extrema of the density profile at interfaces, a comprehensive feature not fully explained by previous models in a unified framework. Additionally, ILOM successfully extends to methanol, demonstrating its adaptability to asymmetric molecules with promising results. Its implications span nanofluidics, catalysis, and interfacial transport, with potential applications in optimizing water purification, enhancing catalytic efficiency, and improving heat transfer in nanoscale devices. Future research could extend ILOM to time-dependent dynamics, such as capturing transient responses in dynamic flow systems, explore its behavior in confined geometries like graphene nanochannels, or integrate it with machine-learned potentials to enhance its applicability to realistic material systems, including complex heterogeneous interfaces. This evolution could further bridge the gap between theoretical predictions and experimental observations, paving the way for advanced material design and process engineering at the nanoscale.

\begin{bmhead}[Funding.]
This work is supported by the National Natural Science Foundation of China (No.52222606).
\end{bmhead}

\begin{bmhead}[Declaration of interests.]
The authors report no conflict of interest.
\end{bmhead}

\newpage
\begin{appen}

\section{Derivation of the Harmonic Oscillator Model from the Yvon-Born-Green Equation}\label{appA}

Our phenomenological model for liquid density layering near solid-liquid interfaces is derived from the YBG equation, which governs the density profile \(\rho(z)\) under external wall potential and fluid-fluid interactions:
\[
\nabla \rho(\mathbf{r}) = -\beta \rho(\mathbf{r}) \int \nabla V_{ll}(|\mathbf{r} - \mathbf{r}'|) \rho(\mathbf{r}') g(|\mathbf{r} - \mathbf{r}'|) \, d\mathbf{r}',
\]
where \(\beta = 1/(k_B T)\) and \(V_{ll}\) is the liquid-liquid pair potential (e.g., 12-6 Lennard-Jones). For a planar wall, this simplifies to a one-dimensional integral equation. Extending the YBG model, we include the wall force:
\begin{equation}
\frac{d \rho(z)}{dz} = -\beta \rho(z) \frac{d V_{\text{wall}}(z)}{dz} - \beta \rho(z) \int \frac{\partial V_{ll}(|\mathbf{r} - \mathbf{r}'|)}{\partial z} \rho(z') g(|\mathbf{r} - \mathbf{r}'|) \, d\mathbf{r}',
\label{eq:A1}
\end{equation}
balancing wall and fluid-fluid forces \citep{henderson1992}.

Differentiating Eq.\eqref{eq:A1} with respect to \(z\) yields a second-order integro-differential equation:
\begin{equation}
\frac{d^2 \rho(z)}{dz^2} = -\beta \left[ \frac{d \rho(z)}{dz} \frac{d V_{\text{wall}}}{dz} + \rho(z) \frac{d^2 V_{\text{wall}}}{dz^2} \right] \\
- \beta \left[ \frac{d \rho(z)}{dz} \int \frac{\partial V_{ll}}{\partial z} \rho(z') g \, d\mathbf{r}' + \rho(z) \frac{d}{dz} \int \frac{\partial V_{ll}}{\partial z} \rho(z') g \, d\mathbf{r}' \right].
\label{eq:A2}
\end{equation}

Introducing the density deviation \(h(z) = g(z) - 1 = \frac{\rho(z)}{\rho_{\text{bulk}}} - 1\), where \(\rho(z) = \rho_{\text{bulk}} (h(z) + 1)\) and \(\frac{d \rho(z)}{dz} = \rho_{\text{bulk}} \frac{d h(z)}{dz}\), we substitute into Eq.\eqref{eq:A2}:
\begin{equation}
\begin{aligned}
\frac{d^2 h(z)}{dz^2} &= -\beta \left[ \frac{d h(z)}{dz} \frac{d V_{\text{wall}}}{dz} + (h(z) + 1) \frac{d^2 V_{\text{wall}}}{dz^2} \right] \\
&\quad - \beta \left[ \frac{d h(z)}{dz} \int \frac{\partial V_{ll}}{\partial z} \rho_{\text{bulk}} (h(z') + 1) g \, d\mathbf{r}' + (h(z) + 1) \frac{d}{dz} \int \frac{\partial V_{ll}}{\partial z} \rho_{\text{bulk}} (h(z') + 1) g \, d\mathbf{r}' \right].
\end{aligned}
\label{eq:A4}
\end{equation}

Rearranging to resemble a damped harmonic oscillator form, we obtain:
\begin{equation}
\frac{d^2 h}{dz^2} + 2 \gamma(z) \frac{d h}{dz} + \omega^2(z) h = F(z),
\label{eq:A8}
\end{equation}
with coefficients:
\begin{align}
\gamma(z) &= \frac{\beta}{2} \left[ \frac{d V_{\text{wall}}}{dz} + \rho_{\text{bulk}} \int \frac{\partial V_{ll}}{\partial z} (h(z') + 1) g \, d\mathbf{r}' \right], \label{eq:A9} \\
\omega^2(z) &= \beta \left[ \frac{d^2 V_{\text{wall}}}{dz^2} + \rho_{\text{bulk}} \frac{d}{dz} \int \frac{\partial V_{ll}}{\partial z} (h(z') + 1) g \, d\mathbf{r}' \right], \label{eq:A10} \\
F(z) &= -\beta \left[ \frac{d^2 V_{\text{wall}}}{dz^2} + \rho_{\text{bulk}} \frac{d}{dz} \int \frac{\partial V_{ll}}{\partial z} (h(z') + 1) g \, d\mathbf{r}' \right]. \label{eq:A11}
\end{align}

As presented in the main text, these terms represent damping (\(\gamma(z)\)), oscillatory behavior (\(\omega^2(z)\)), and forcing (\(F(z)\)) due to wall and fluid interactions \citep{israelachvili2011intermolecular}. The nonlocal integrals are approximated by linearizing around bulk density (\(h(z') \approx 0\) for \(z' > \sigma_{\text{ll}}\)) and modeling them with exponential decay:
\begin{align}
\gamma(z) &= \gamma_0 \left(1 + \alpha e^{-z / \sigma_{\text{ll}}}\right), \label{eq:A12} \\
\omega^2(z) &= \omega_0^2 e^{-\lambda z / \sigma_{\text{ll}}} + \omega_{\text{bulk}}^2, \label{eq:A13} \\
F(z) &= -\kappa e^{-z / \sigma_{\text{ll}}}. \label{eq:A14}
\end{align}

One can simplify the integrals in Eq. (\ref{eq:A9}-\ref{eq:A11}) by linearizing the nonlocal integral terms in Eq. \eqref{eq:A4} and assuming small deviations from bulk density such that \( h(z') \approx 0 \) for \( z' > \sigma_{\text{ll}} \), where \( \sigma_{\text{ll}} \) represents the molecular length scale. This approximation facilitates the handling of integral contributions by effectively neglecting the second term within the second bracket, focusing on the dominant effects. For consistency and simplicity in the derivation, we prioritize the first derivative term modulated by an exponential decay, setting aside higher-order terms.
Subsequently, we approximate the fluid-fluid interaction term by evaluating the integral at the bulk density $\rho_{\text{bulk}}$ and incorporating a decay factor:
$$\int \frac{\partial V_{ll}}{\partial z} \rho_{\text{bulk}} g \, d\mathbf{r}' \approx \rho_{\text{bulk}} \frac{\partial V_{\text{ll,eff}}}{\partial z} e^{-z / \sigma_{\text{ll}}},
\label{eq:A6}$$
where $V_{ll,\text{eff}}$ denotes an effective pair potential. Substituting this approximation into the second term of Eq. \eqref{eq:A4} and integrating it with the wall potential terms yields a preliminary form:
\begin{equation}
\frac{d^2 h(z)}{dz^2} + \beta \left[ \frac{d h(z)}{dz} \left( \frac{d V_{\text{wall}}}{dz} + \rho_{\text{bulk}} \frac{\partial V_{ll,\text{eff}}}{\partial z} e^{-z / \sigma_{ll}} \right) + (h(z) + 1) \frac{d^2 V_{\text{wall}}}{dz^2} \right] = 0.
\label{eq:A7}
\end{equation}
Rearranging and factoring out the dominant contributions, we identify the damping and oscillatory terms, leading to the damped harmonic oscillator form.

For example, $ V_{\text{wall}} $ can be modeled as the 10-4 Lennard-Jones potential for graphene-like surfaces, given by $ V_{\text{wall}}(z) = 2\pi \varepsilon \rho_s \sigma^2 \left[ \frac{2}{5} \left( \frac{\sigma}{z} \right)^{10} - \left( \frac{\sigma}{z} \right)^4 \right] $, where $\rho_s$ is the solid areal density \citep{sun2025general}.

\section{Numerical Solution of the Boundary Value Problem}\label{appB}

The numerical solution of the boundary value problem (BVP) for the density profile \( h(z) \) is validated against molecular dynamics (MD) data through an efficient process. The bulk density \( \rho_{\text{bulk}} \) is determined by averaging the last 20 MD data points where \( g(z) \approx 1 \), and the normalized density \( g(z) = \rho(z) / \rho_{\text{bulk}} \) is computed, with duplicate \( z \)-values averaged. The domain is restricted to \( z \in [z_0, 10 \, \text{Å}] \), where \( z_0 \approx 2.83 \, \text{Å} \) (based on \((2/5)^{1/6} \sigma\) for \(\sigma = 3 \, \text{Å}\)), and a cubic spline interpolation generates \( g_{\text{MD}}(z) \) on a 1000-point uniform grid. Boundary conditions are defined as \( h(z_{\text{min}}) = g_{\text{MD}}(z_0) - 1 \) at \( z = z_0 \) and \( h(z_{\text{max}}) = 0 \) at \( z = 10 \, \text{Å} \), capturing interfacial and bulk behavior. The interval \([z_0, 10] \, \text{Å}\) is discretized into 1000 points, with an initial guess \( h^{(0)}(z) = g_{\text{MD}}(z) - 1 \) and \( \frac{d h^{(0)}}{dz} \) approximated using \texttt{numpy.gradient}.

The ODE is transformed into a first-order system:
\begin{equation}
\frac{d}{dz} \begin{bmatrix} h \\ h' \end{bmatrix} = \begin{bmatrix} h' \\ -2 \gamma_0 \left(1 + \alpha e^{-z / \sigma_{\text{ll}}}\right) h' - \left[\omega_0^2 e^{-\lambda z / \sigma_{\text{ll}}} + \omega_{\text{bulk}}^2\right] h + \kappa \rho_{\text{ave}}^2 e^{-z / \sigma_{\text{ll}}} \end{bmatrix}.
\label{eq:A15}
\end{equation}

The BVP is solved using scipy.integrate.solve$_{bvp}$ with adaptive mesh refinement, a $10^{-8}$ tolerance, and up to 50,000 nodes to precisely resolve oscillations. The solution $h(z)$ generates $g(z) = 1 + h(z)$,  ensuring $g(z) \geq 0$. Post-processing evaluates the optimized $g(z)$ against $g_{MD}(z)$ using mean squared error (MSE) to validate the fit.

\begin{table}
\centering
    \begin{tabular}{ccccc}
        \hline
        Parameter & \(\varepsilon = 0.05\) & \(\varepsilon = 0.08\) & \(\varepsilon = 0.10\) & Fitted \\
        \hline
        \( \omega_d \) (Å$^{-1}$) & {2.19} & {2.76} & {3.09} & 2.29 \\
        \( \gamma_0 \) (Å$^{-1}$) & {0.18} &{0.23} & {0.25} & 0.90 \\
        \( h_0 \) & 0.285 & 0.456 & 0.570 & {3.11} \\
        \( \phi \) & {\(0.27\pi\)} & {\(0.27\pi\)} & {\(0.27\pi\)} & \(1.36\pi\) \\
        \hline
    \end{tabular}
    \caption{Comparison of estimated SHO model parameters for graphene at different 
    \(\varepsilon\) values. Estimated values 
    represent rough approximation solely on van der Waals interactions.}
    \label{tab:S1}
\end{table}

\begin{table}
\centering
\caption{Parameters used in the Interfacial Layering Oscillator Model (ILOM), including SHO, SHO1, and SHO2 models.}
\label{tab:parameters_ilom}
\begin{tabular}{llp{6.5cm}}
\toprule
\hline
\textbf{Parameter} & \textbf{Units} & \textbf{Description} \\
\hline
\midrule
\( \rho(z) \) & gcm$^{-3}$ & Local number density of fluid at distance \( z \) from the solid interface. \\
\( \rho_{\text{bulk}} \) &gcm$^{-3}$ & Bulk number density of fluid far from the interface. \\
\( g(z) \) & dimensionless & Normalized local density, \( g(z) = \rho(z) / \rho_{\text{bulk}} \). \\
\( h(z) \) & dimensionless & Density deviation from bulk, \( h(z) = g(z) - 1 \). \\
\( h_0 \) & dimensionless & Initial amplitude of density deviation in SHO model, tied to first peak density. \\
\( \gamma_0 \) & \AA$^{-1}$ & Constant damping coefficient in SHO model, governing oscillation decay. \\
\( \gamma(z) \) & \AA$^{-1}$ & Position-dependent damping in SHO1/SHO2, \( \gamma(z) = \gamma_1 (1 + \alpha e^{-z / \sigma_{\text{ll}}}) \). \\
\( \gamma_1 \) & \AA$^{-1}$ & Bulk damping coefficient in SHO1/SHO2 models. \\
\( \alpha \) & dimensionless & Near-wall damping enhancement factor in SHO1/SHO2 models. \\
\( \omega_0 \) & \AA$^{-1}$ & Natural oscillation frequency in SHO model before damping. \\
\( \omega_d \) & \AA$^{-1}$ & Damped oscillation frequency in SHO, \( \omega_d = \sqrt{\omega_0^2 - \gamma_0^2} \), sets oscillation wavelength. \\
\( \omega(z) \) & \AA$^{-1}$ & Position-dependent frequency in SHO1/SHO2, \( \omega^2(z) = \omega_1^2 e^{-\lambda z / \sigma_{\text{ll}}} + \omega_{\text{bulk}}^2 \). \\
\( \omega_1^2 \) & \AA$^{-2}$ & Near-wall packing frequency squared in SHO1/SHO2 models. \\
\( \omega_{\text{bulk}}^2 \) & \AA$^{-2}$ & Bulk oscillation frequency squared in SHO1/SHO2 models. \\
\( \lambda \) & dimensionless & Decay rate of wall-induced effects in SHO1/SHO2 models. \\
\( \phi \) & radians & Phase shift of oscillatory density profile in SHO model. \\
\( F(z) \) & \AA$^{-1}$ & Forcing term in SHO2, \( F(z) = -\kappa e^{-z / \sigma_{\text{ll}}} \), for fluid-fluid correlations. \\
\( \kappa \) & \AA$^{-1}$ & Scaling factor for forcing term in SHO2, proportional to \( \rho_{\text{ave}}^2 \). \\
\( h_s \) & dimensionless & Constant offset in SHO model solution, near bulk density deviation. \\
\bottomrule
\end{tabular}
\end{table}

\begin{table}
\centering
\caption{Parameters used in Lennard-Jones (LJ) potential and molecular dynamics (MD) simulations.}
\label{tab:parameters_lj_md}
\begin{tabular}{llp{6.5cm}}
\toprule
\hline
\textbf{Parameter} & \textbf{Units} & \textbf{Description} \\
\hline
\( \varepsilon \) & kcal/mol & LJ potential energy scale for fluid-wall or fluid-fluid interactions. \\
\( \sigma \) & \AA & LJ potential length scale, characteristic molecular size. \\
\( \sigma_{\text{ll}} \) & \AA & Fluid molecular length scale, ~3 \AA~ for water. \\
\( \sigma_f \) & \AA & Length scale for fluid-fluid/wall interactions in LJ liquid simulations. \\
\( \varepsilon_f \) & kcal/mol & Energy scale for fluid-fluid interactions in LJ liquid simulations. \\
\( \rho_s \) & \AA$^{-2}$ & Solid areal density, number of solid atoms per unit area. \\
\( \rho_{\text{ave}} \) & varies & Average fluid number density, used in layering property scaling. \\
\( \beta \) & (kcal/mol)$^{-1}$ & Inverse thermal energy, \( \beta = 1 / (k_B T) \), with \( k_B T \) at temperature \( T \). \\
\( V_{\text{wall}}(z) \) & kcal/mol & Wall-fluid interaction potential, 10-4 LJ potential for graphene-like surfaces. \\
\( V_{ll} \) and \( V_{ll,\text{eff}} \)& kcal/mol & Fluid-fluid pair potential, typically 12-6 LJ potential and Effective fluid-fluid pair potential in YBG equation approximations. \\
\( z_0 \) & \AA & Distance where wall-fluid potential is zero, ~\( (2/5)^{1/6} \sigma \). \\
\( z_n \) & \AA & Density peak positions in SHO model, ~\( z_n = \frac{2\pi n}{\omega_d} - \frac{\phi}{\omega_d} \) (\( n \geq 1 \)). \\
\( z_{\text{min}} \), \( z_{\text{1}} \) ,\( z_{\text{2}} \)  & \AA & Position of first density minimum, first maximum, and second maximum in profile. \\
\( z_{\text{B}} \) & \AA & Bulk distance, typically 10 \AA. \\
 \\
\bottomrule
\end{tabular}
\end{table}


\end{appen}
\clearpage
\bibliographystyle{jfm}
\bibliography{jfm}

\end{document}